\documentclass[twocolumn,showpacs,preprintnumbers,amsmath,amssymb,floatfix]{revtex4}

\usepackage{graphicx}
\usepackage{dcolumn}
\usepackage{bm}

\newcommand{\beq}{\begin{equation}}
\newcommand{\eeq}{\end{equation}}
\newcommand{\bey}{\begin{eqnarray}}
\newcommand{\eey}{\end{eqnarray}}

\begin{document}

\title{Charged gravastars admitting conformal motion}

\author{A. A. Usmani}
\email{anisul@iucaa.ernet.in} \affiliation{Department of Physics,
Aligarh Muslim University, Aligarh 202 002, Uttar Pradesh, India}

\author{Farook Rahaman}
\email{farook\_rahaman@yahoo.com} \affiliation{Department of
Mathematics, Jadavpur University, Kolkata 700 032, West Bengal,
India}

\author{Saibal Ray}
\email{saibal@iucaa.ernet.in} \affiliation{Department of Physics,
Government College of Engineering and Ceramic Technology, Kolkata
700 010, West Bengal, India}

\author{K. K. Nandi}
\email{kamalnandi1952@yahoo.co.in} \affiliation{Department of
Mathematics, North Bengal University, Siliguri 734013, West
Bengal, India}

\author{Peter K. F. Kuhfittig}
\email{kuhfitti@msoe.edu}
\affiliation{Department of Mathematics,
Milwaukee School of Engineering, Milwaukee, Wisconsin 53202-3109,
USA}

\author{Sk. A. Rakib}
\affiliation{Department of Mathematics, Jadavpur University,
Kolkata 700 032, West Bengal, India}

\author{Z. Hasan}
\affiliation{Department of Physics and Astrophysics, University of
Delhi, Delhi- 110007, India }


\begin{abstract}\noindent
We propose a new model of a {\it gravastar} admitting conformal
motion.  While retaining the framework of the Mazur-Mottola
model, the gravastar is assumed to be internally charged, with an
exterior defined by a Reissner-Nordstr{\"o}m rather than a
Schwarzschild line element. The solutions obtained involve (i) the
interior region, (ii) the shell, and (iii) the exterior region of
the sphere. Of these three cases the first case is of primary
interest since the total gravitational mass vanishes for vanishing
charge and turns the total gravitational mass into an {\it
electromagnetic mass } under certain conditions. This suggests
that the interior de Sitter vacuum of a charged gravastar is
essentially an electromagnetic mass model that must generate the
gravitational mass. We have also analyzed various other aspects
such as the stress energy tensor in the thin shell and the
entropy of the system.
\end{abstract}


\maketitle

\section{Introduction}\noindent
By extending the concept of Bose-Einstein condensate to
gravitational systems, Mazur and Mottola
\cite{Mazur2001,Mazur2004} have proposed a new solution  for the
endpoint of a gravitational collapse in the form of cold, dark, compact
objects known as {\it gravastars}.  These contain an isotropic
de Sitter vacuum  in the interior, while the exterior is defined by
a Schwarzschild geometry, separated by a thin shell of stiff matter
of arbitrary total mass $M$. This model implies that the space of a
gravastar has three different regions with different equations of state (EOS)
\cite{Visser2004,Bilic2005,Cattoen2005,Carter2005,Lobo2006,DeBenedictis2006,
Lobo2007,Horvat2007,Cecilia2007,Rocha2008,Horvat2008,Nandi2009,Turimov2009},
as defined as follows:

I. Interior: $0 \leq r < r_1 $, ~~~$ p = -\rho $,

II. Shell: $ r_1 < r < r_2 $, ~~~$ p = +\rho $,

III. Exterior: $ r_2 < r $, ~~~$ p = \rho =0. $\newline
Here $r_2-r_1=\delta$ is the thickness of the shell. The
presence of matter on the thin shell is required to achieve the
crucial stability of such systems under expansion by exerting
an inward force to balance the repulsion from within.

Building on this background, we propose a new model of a gravastar
admitting conformal motion by assuming a charged interior but with
an exterior defined by a Reissner-Nordstr{\"o}m line element
instead of Schwarzschild. The basic motivation for this model
is that, in general, compact stars tend to have a net charge on
the surface
~\cite{Stettner1973,Whitman1981,Ray2003,Ghezzi2005,Ray2007a}.
This is an essential consideration in the study of the stability
of a fluid sphere; in fact, it has been argued
~\cite{Stettner1973,Whitman1981} that a spherical fluid distribution
of uniform density with a net surface charge is more stable than a
surface without charge. According to de Felice
et al.~\cite{Felice1995}, the inclusion of charge inhibits
the growth of spacetime curvature and which therefore plays a
key role in avoiding  singularities. It has also been argued
that gravitational collapse can be averted in the presence of
a charge since the  gravitational attraction may be
counter-balanced by the electrical repulsion (in addition to
the pressure gradient~\cite{Sharma2001,Ivanov2002}.)

In searching for a natural relationship between geometry and matter
for such stars through the Einstein field equations, we take
into account the well-known inheritance symmetry. This symmetry is
contained in the set of conformal killing vectors (CKV)
\begin{equation}
L_\xi g_{ij} = \psi g_{ij}.  \label{ckv}
\end{equation}
Here $L$ is the Lie derivative operator and $\psi$ is the
conformal factor. It is supposed that the vector $\xi$ generates
the conformal symmetry and the metric $g$ is conformally mapped
onto itself along $\xi$. Neither $\xi$ nor $\psi$ need to be
static, even in the case of a static metric
~\cite{Bohmer2008a,Bohmer2008b}. Due to this and other
properties, CKVs have provided a deeper insight into the
spacetime geometry connected to the astrophysical and
cosmological realm~\cite{Mars1994,Maartens1996,Mak2004a,
Harko2004, Mak2004b,Neves2006,Rahaman2010a,Rahaman2010b}.

It has been shown by Ray et al. \cite{Ray2008} that, under
the conformal killing-vectors approach, charged fluid
spheres provide \emph{electromagnetic mass (EMM)} models in which
gravitational mass and other physical parameters originate solely
from the electromagnetic field. So one of the motivations in the
present investigation is to include CKVs and see to what extent
conformal motion admits EMM models along with other relevant physical
features. This paper is organized as follows: in Sec.~II the
Einstein-Maxwell field equations are provided, along with the
CKVs for a charged gravastar.  The solutions are obtained in
Sec. III for a charged gravastar with conformal motion in connection
with (A) the interior region, (B) the  shell, and (C) the exterior
region of the sphere. Secs. IV, V, and VI deal with the stress energy
tensor in the thin shell, entropy within the shell, and unknown
constants, respectively.  In Sec. VII we conclude.

\section{The Einstein-Maxwell Field Equations}
\noindent
The Einstein field equations for the case of a charged perfect
fluid source are
\begin{equation}
{G^{i}}_{j} = {R^{i}}_{j} - \frac{1}{2}{{g^{i}}_{j}} R = -\kappa
\left[{{T^{i}}_{j}}^{(m)} + {{T^{i}}_{j}}^{(em)}\right],
\end{equation}
where the energy-momentum tensor components for the matter
source and electromagnetic field, respectively, are given by
\begin{equation}
{{T^{i}}_{j}}^{(m)} = (\rho + p) u^{i}u_{j} + p{g^{i}}_{j},
\end{equation}
\begin{equation}
{{T^{i}}_{j}}^{(em)} = -\frac{1}{4\pi}\left[F_{jk}F^{ik} -
\frac{1}{4} {g^{i}}_{j}F_{kl} F^{kl}\right].
\end{equation}
Here $\rho$, $p$ and $u^{i}$ are the matter-energy
density, the fluid pressure, and the velocity four-vector of a
fluid element (with $u_{i}u^{i}=1$), respectively. The corresponding
Maxwell electromagnetic field equations are
\begin{equation}
{[{(- g)}^{1/2} F^{ij}],}_{j} = 4\pi J^{i}{(- g)}^{1/2},
\end{equation}
\begin{equation}
F_{[ij,k]} = 0,\label{max2}
\end{equation}
where the electromagnetic field tensor $F_{ij}$ is related to the
electromagnetic potentials through the relation $ F_{ij} = A_{i,j}
- A_{j,i} $ and is equivalent to Eq.~(\ref{max2}). In the
above equations, $J^{i}$ is the current four-vector satisfying
$J^{i} = \sigma u^{i}$, where $\sigma$ is the charge density and
$ \kappa = 8 \pi $, using relativistic units $G = c = 1$. Here
and in what follows a comma denotes the partial derivative with
respect to the coordinates.

Next, given the static spherically symmetric space-time
\begin{equation}
ds^2=-e^{\nu(r)}dt^2+ e^{\lambda(r)} dr^2+r^2(
d\theta^2+\text{sin}^2\theta\, d\phi^2),
\end{equation}
the Einstein-Maxwell field equations may be written as
\begin{equation}e^{-\lambda}
\left[\frac{\lambda^\prime}{r} - \frac{1}{r^2}\right
]+\frac{1}{r^2}= 8\pi \rho + E^2, \label{ein1}
\end{equation}
\begin{equation}e^{-\lambda}
\left[\frac{1}{r^2}+\frac{\nu^\prime}{r} \right]-\frac{1}{r^2}=
8\pi p - E^2,\label{ein2}
\end{equation}
\begin{equation}\frac{1}{2} e^{-\lambda}
\left[\frac{1}{2}(\nu^\prime)^2+ \nu^{\prime\prime}
-\frac{1}{2}\lambda^\prime\nu^\prime + \frac{1}{r}({\nu^\prime-
\lambda^\prime}) \right] = 8\pi p + E^2, \label{ein3}
\end{equation}
and
\begin{equation}
[r^2E]^\prime = 4\pi r^2 \sigma e^{\lambda/2}.
 \label{max3}
\end{equation}
Equation (\ref{max3}) may be expressed for the electric
field $E$ in the following equivalent form:
\begin{equation} E(r) = \frac{1}{r^2}\int_0^r 4\pi r^2 \sigma
e^{\lambda/2}dr = \frac{q(r)}{r^2}, \label{max4}
\end{equation}
where $q(r)$ is the total charge of the sphere.

\section{The Charged Gravastar with conformal motion}
\noindent
Equation~(\ref{ckv}) implies the following:
\begin{equation}
L_\xi g_{ik} =\xi_{i;k}+ \xi_{k;i} = \psi g_{ik} \label{Eq9}
\end{equation}
with $\xi_i = g_{ik}\xi^k$. Here $1$ and $4$ stand for the
spatial and temporal coordinates $r$ and $t$,
respectively.

Equations (\ref{Eq9}) yield the following expressions \cite{Ray2008}:
\begin{eqnarray}
&\xi^1 \nu^\prime =\psi,\nonumber \\
&\xi^4  = C_1 \nonumber \\
&\xi^1  = \frac{\psi r}{2},\nonumber \\
&\xi^1 \lambda ^\prime + 2 \xi^1 _{,1} =\psi, \nonumber
\end{eqnarray}
which imply
\begin{eqnarray}
e^\nu  &=& C_2^2 r^2, \label{nu}\\ e^\lambda  &=& \left[\frac
{C_3} {\psi}\right]^2,  \label{lam} \\ \xi^i &=& C_1 \delta_4^i +
\left[\frac{\psi r}{2}\right]\delta_1^i, \label{Eq12}
\end{eqnarray}
where $C_1$, $C_2$, and $C_3$ are integration constants.

Given solutions (\ref{nu}) and (\ref{lam}), Eqs. (\ref{ein1}),
(\ref{ein2}), and (\ref{ein3}) take the following form
\cite{Ray2008}:
\begin{eqnarray}
\frac{1}{r^2}\left[1 - \frac{\psi^2}{C_3^2}
\right]-\frac{2\psi\psi^\prime}{rC_3^2}&=& 8\pi \rho + E^2, \label{Eq13} \\
\frac{1}{r^2}\left[1 - \frac{3\psi^2}{C_3^2}
\right]&=& - 8\pi p + E^2,   \label{Eq14}\\
\left[\frac{\psi^2}{C_3^2r^2}
\right]+\frac{2\psi\psi^\prime}{rC_3^2} &=& 8\pi p + E^2. \label{Eq15}
\end{eqnarray}

From the above equations, one may easily obtain values for $E$,
$\rho$ and $p$ \cite{Ray2008}:
\begin{eqnarray}
E^2 &=& \frac{1}{2} \left[ \frac{1}{r^2}\left(1 -
\frac{2\psi^2}{C_3^2}
\right)+\frac{2\psi\psi^\prime}{rC_3^2}\right], \label{charge}
\\
8\pi \rho  &=&  \frac{1}{2r^2} -\frac{3\psi\psi^\prime}{rC_3^2},
\label{density}
\\
8\pi p &=& \frac{\psi\psi^\prime}{rC_3^2} - \frac{1}{2r^2} \left[1
- \frac{4\psi^2}{C_3^2} \right].\label{pressure}
\end{eqnarray}

\subsection{Interior region of the charged gravastar}
\noindent
Observe next that Eqs.~(\ref{density}) and (\ref{pressure})
provide an essential relationship between the metric potentials
and the physical parameters $\rho$ and $p$:
\begin{equation}
\frac{2\psi}{r^2C_3^2}(\psi-r\psi^\prime) = 8 \pi r(\rho +
p).\label{key1}
\end{equation}
So given the {\it ansatz} $\rho + p = 0$, it follows
from the Eq.~(\ref{key1}) that the value of $\psi$ turns
out to be either $\psi=0$ or $\psi = \psi_0 r$, where
$\psi_0$ is a dimensionless integration constant.  This
leads to the following exact analytical forms for all of
the parameters:
\begin{eqnarray}
8\pi \rho = \frac{1}{2r^2}-3\widetilde{\psi}_0^2=-8\pi p,
\label{Eq20}\\
E^2 = \frac{1}{2r^2}, \label{Eq21}\\
e^\nu = e^{-\lambda}= \widetilde{\psi}_0^2 r^2, \label{Eq23}   \\
\sigma=\frac{\widetilde{\psi}_0}{4\pi \sqrt{2} r}. \label{Eq24}
\end{eqnarray}
Here $\widetilde{\psi}_0=\psi_0/C_3$ is a constant, which has
the inverse dimension of $r$.  The reason is that under the
condition $\rho + p = 0$, $\widetilde{\psi}_0=C_2$, so that
$C_2C_3 = \psi_0$.


It is clear from Eq.~(\ref{Eq20}) that $1/2r^2 -
3\widetilde{\psi}_0^2>0$ (or $\widetilde{\psi}_0^2 <1/6r^2$)
illustrates a case of positive density and negative pressure,
resulting in an outward push from the interior region, which is
consistent with the physics of a gravastar. On the other hand,
$1/2r^2-3\widetilde{\psi}_0^2<0$ (or
$\widetilde{\psi}_0^2<1/6r^2$) represents a collapsing case with
negative density and positive pressure, which is not the subject
of concern here. Thus for the purpose of gravastar physics, our
above solutions are assumed to obey the condition
$0< \widetilde{\psi}_0^2<1/6r^2$.

For $\widetilde{\psi}_0^2=0$, we find that both $p$ and $\rho$ are
inversely proportional to $r^2$ but with opposite signs for their
proportionality constants. A nonzero value of
$\widetilde{\psi}_0^2$ leads to a translational shift of this
form with magnitude $3\widetilde{\psi}_0^2$, as can be seen from from
Eq.~(\ref{Eq20}). The electric field $E$ is found to be inversely
proportional to $r$ and is independent of $\widetilde{\psi}_0$.
On the other hand, $\sigma$ is inversely proportional to $r$. Its
value is zero for $\widetilde{\psi}_0=0$, which suggests that the
above power law behavior of $p$ and $\rho$ is independent of
$\sigma$.  Both $e^\nu$ and $e^{-\lambda}$ are proportional to
$r^2$ and hence equal with equal proportionality constants.

The active gravitational mass $M(r)$, by virtue of the field
Eq.~(\ref{ein1}), may be expressed in the following form:
\begin{equation}
M(r) = 4 \pi \int_{0}^{r}\left[\rho +  \frac{E^2}{8 \pi}\right]
r^2 dr = \frac{1}{2}r(1-\widetilde{\psi}_0^2 r^2).\label{mass}
\end{equation}
Here the pressure and density fail to be regular at the origin,
but the effective gravitational mass is always positive since
$\widetilde{\psi}_0^2<1/6r^2$ and will vanish as
$r \rightarrow 0$.  In other words, the expression for $M(r)$
does not lead to a singularity.

Let us now match the interior solution to the exterior
Reissner-Nordstr\"{o}m solution at the boundary in the
customary manner.  To do so, we have to keep in mind that
instead of a solid sphere, we are dealing here with a
bubble-like hollow sphere of radius $r_2 = r_1+\delta$, so
that for the limit $\delta \rightarrow 0$, we have the de
Sitter spherical void of radius $r_2 \rightarrow r_1$.
(According to Ref. \cite{Mazur2001}, $\delta$ does not exceed
Planck length by more than a few orders of magnitude.)  For
convenience of notation, let us denote the radius $r_2$ by $a$,
the radius of the junction surface.  Then following Ray et al.
\cite{Ray2008}, the total gravitational mass $m(r=a)$, which
is obtained after matching the solution interior to $r=a$ to
the exterior Reissner-Nordstr{\"o}m solution at the boundary,
can be expressed as
\begin{equation}
m(a) =M(a) + \frac{q(a)^2}{2a} = \frac{1}{2\sqrt 2}(3
- 8\widetilde{\psi}_0^2q^2)q, \label{totmass}
\end{equation}
where $M(a)$ is the total active gravitational mass and
$q(a)^2/2a$ is the mass equivalence of the electromagnetic
field. Observe from Eq.~(\ref{totmass}) that the vanishing
charge $q$ turns the total gravitational mass $m$ into an
EMM under the constraint $\widetilde{\psi}_0<\sqrt
3/(2\sqrt{2})q$.

It is interesting to note that the mathematical
expressions and physics of the interior region resemble
the EMM model of Ray et al.~\cite{Ray2008}. The apparent
reason for this is the use of the Reissner-Nordstr{\"o}m line
element. In other words, the interior de Sitter void
($p=-\rho$) in a charged gravastar is the same as in the case
already addressed by Ray et al. \cite{Ray2008}. This implies
that the interior de Sitter void of a charged gravastar
must, in analogous fashion, generate the gravitational mass.
This particular feature is a new one and was not obtainable
in the non-charged case of Mazur and Mottola
~\cite{Mazur2001,Mazur2004}. This mass provides the attractive
force resulting from the collapse of the sphere and
counter-balances the repulsive force due to electromagnetic field.

It is worth noting that the equation of state $p=-\rho$ (known
in the literature as a false vacuum, degenerate vacuum, or
$\rho$-vacuum ~\cite{Davies1984,Blome1984,Hogan1984,Kaiser1984})
represents a repulsive pressure which in the context of an
accelerating Universe may be related to the {\it
$\Lambda$-dark energy}, an agent responsible for the second phase
of the inflation
~\cite{Riess1998,Perlmutter1999,Ray2007b,Usmani2008,Frieman2008}.
So the {\it charged gravastar} seems to be connected to
the {\it dark star}~\cite{Chan2009a,Chan2009b,Lobo2008}.

\subsection{Shell of the charged gravastar}
\noindent
Using the EOS $p=\rho$, we get the solution
\begin{equation}
\psi^2 = \frac{C_2^2}{2} - \frac{\psi_1}{ r}, \label{Eq26}
\end{equation}
where $\psi_1 > 0$ is an integration constant.

Other parameters are
\begin{eqnarray}\label{E:perfect}
8\pi\rho
&=&\frac{1}{2r^2}\left(1-\frac{3\widetilde{\psi}_1}{r}\right)=8\pi
p, \label{Eq27}\\ E^2 &=& \frac{1}{2 r^2}-8\pi \rho,
\label{Eq28}\\ e^\nu  &=& C_1^2 r^2, \label{Eq29}\\ e^{-\lambda}
&=& 1/2 - \widetilde{\psi}_1/r, \label{Eq30}\\ \sigma  &=&
\frac{\sqrt{3} \widetilde{\psi}_1}{16\pi r^3}\sqrt{\frac{r}{
\widetilde{\psi}_1}-2}, \label{Eq31}
\end{eqnarray}
where $\widetilde{\psi}_1= \psi_1/C_2^2$ has the same
dimension as $r$. Although the electric field $E$ is inversely
proportional to $r$, it depends on the integration constant
$\widetilde{\psi}_1$, unlike the previous case.  Eq.~(\ref{Eq31})
suggests that the requirement of a real-valued $\sigma$ may
only be achieved with the condition $\widetilde{\psi}_1< r/2$.
Combining this with the previous condition, $\psi_1>0$ (or
$\widetilde{\psi}_1>0$), we observe that the above solutions for
the shell of the gravastar are valid within the range
$0<\widetilde{\psi}_1 < r/2$. It is obvious from Eq.~(\ref{Eq27})
that the EOS $p=\rho=0$ for the exterior de Sitter region
corresponds to $\widetilde{\psi}_1=r/3$, which is within the upper
limit of the above condition $\widetilde{\psi}_1<r/2$.

The proper thickness of the shell is obtained next:
\begin{eqnarray}
\ell &=& \int _{r_1}^{r_2} \sqrt{e^\lambda } dr  \nonumber \\&=&
\sqrt{2}\left[R +\widetilde{\psi}_1\ln \left(R +r -
\widetilde{\psi}_1\right)
            \right]_{a}^{a+\epsilon},\label{Eq32}
\end{eqnarray}
where $ R= r\sqrt{1-2\widetilde{\psi}_1/r}$. Thus a real value for
the thickness confirms that the integration constant
$\widetilde{\psi}_1$ must be less than $r/2$. This condition also
suggests a real $\sigma$.

Finally, using the symbol $\widetilde{E}$ for the energy, we get
within the shell
\begin{equation}
\widetilde{E}  = 4\pi \int _{r_1}^{r_2} \left[ \rho +
\frac{E^2}{8\pi}\right]r^2 dr = \frac{1}{4}[r_2 -r_1].
\label{Eq33}
\end{equation}
Thus $\widetilde{E}$ is exactly proportional to the
coordinate thickness of the shell, (as opposed to the proper
thickness).

\subsection{Exterior region of the charged gravastar}\noindent
For the exterior region ($p = \rho = 0$), the
Reissner-Nordstr\"{o}m space-time is
\begin{equation}
ds^2=-f(r) dt^2+ \frac{ dr^2}{f(r)}+r^2( d\theta^2
  +\text{sin}^2\theta d\phi^2),
\label{Eq35}
\end{equation}
where
\begin{equation}
f(r)=1-\frac{2M}{r} + \frac{Q^2}{r^2}. \label{Eq36}
\end{equation}

\section{The stress energy tensor in the thin shell}\noindent
The second fundamental forms associated with the two
sides of a thin shell (junction surface) are
\cite{Stettner1973,Whitman1981}
\begin{equation}
K_{ij}^\pm = \left. - n_\nu^\pm\ \left[ \frac{\partial^2X^\nu} {\partial
\xi^i\partial \xi^j } +
 \Gamma_{\alpha\beta}^\nu \frac{\partial X^\alpha}{\partial \xi^i}
 \frac{\partial X^\beta}{\partial \xi^j }\right]\right |_S, \label{Eq37}
 \end{equation}
where $ n_\nu^\pm\ $ are the unit normals to $S$:
\begin{equation}
n_\nu^\pm =\pm\left | g^{\alpha\beta}\frac{\partial f}{\partial
X^\alpha} \frac{\partial f}{\partial X^\beta}\right |^{-\frac{1}{2}}
\frac{\partial f} {\partial X^\nu} \label{Eq38}
 \end{equation}
with $ n^\mu n_\mu = 1 $. Here $\xi^i$ are the intrinsic
coordinates on the shell, $f =0 $ is the parametric equation
of the shell $S$, and $-$ and $+$ corresponds to the interior
and exterior regions, respectively.

Using the Lanczos equations
\cite{Israel1966,Rahaman2009a,Rahaman2009b,Rahaman2010,Usmani2010},
one can find the surface stress energy $\Sigma$ and the
surface tangential pressures $ p_\theta = p_\phi \equiv p_t $:
\begin{eqnarray}
\Sigma &=& - \frac{1}{4\pi a}\left[ \sqrt{e^{-\lambda}}\right]_-^+
\nonumber \\&=& - \frac{1}{4\pi a}\left[ \sqrt{
1-\frac{2M}{a} + \frac{Q^2}{a^2}} -{\widetilde{\psi}_0
a}\right], \label{Eq39}
\end{eqnarray}
\begin{eqnarray}\label{E:lateral}
p_t &=&\frac{1}{8\pi a}\left[\left(1+\frac{a\nu'}{2}
   \right)\sqrt{e^{-\lambda}}\right]_-^+
   \nonumber\\ &=&
\frac{1}{4\pi a}\left[\frac{ 1 -M/a}
{2\sqrt{ 1-2M/a +Q^2/a^2}} -{\widetilde{\psi}_0 a}\right].
\label{Eq40}
\end{eqnarray}

The surface mass $M_{shell}$ of this thin shell may be defined
as
\begin{eqnarray}
M_{shell} &=& 4 \pi a^2 \Sigma \nonumber \\&=& -  a \left[
\sqrt{ 1-\frac{2M}{a} + \frac{Q^2}{a^2}}
-{\widetilde{\psi}_0 a}\right]. \label{Eq41}
\end{eqnarray}
Here $M$ can be interpreted as the total mass of the
Reissner-Nordstr\"{o}m gravastar. It takes the following form:
\begin{equation}
M= \frac{1}{ 2 a}  \left[ a^2 +Q^2 + 2a^2
{\widetilde{\psi}_0 } M_{shell} -
  M_{shell}^2 - {\widetilde{\psi}_0^2 a^4} \right]. \label{Eq42}
\end{equation}
Now let $p_\theta = p_\phi  = -v_{\theta} = -v_{\phi}= -v $,
where $v_{\theta}$ and $v_{\phi}$ are the surface tensions.  Then
if Eqs. (\ref{Eq39}) and (\ref{Eq40}) are substituted in the form
\begin{equation}
v =  \omega(a) \Sigma, \label{E:omega}
\end{equation}
the EOS becomes
\begin{multline}
\omega(a) =\\\frac{(1/2\widetilde{\psi}_0 a)\left(1-M/a\right )
-{\sqrt{ 1-2M/a+Q^2/a^2}} }
{(1/\widetilde{\psi}_0 a)\left( 1-2M/a +Q^2/a^2\right )
-\sqrt{ 1-2M/a+Q^2/a^2} }. \label{Eq44}
\end{multline}
With the requirement of a positive density and positive pressure,
the equation of state parameter $\omega(a)$ is always positive.
The position of the thin shell (junction surface) plays a crucial
role: if $a$ is sufficiently large, then $\omega(a)\approx 1$.
For some value of $a$ in Eq.~(\ref{E:lateral}), we may get $p_t =0$,
yielding a dust shell.

\section{Entropy within the shell}\noindent
Following Mazur and Mottola \cite{Mazur2001,Mazur2004}, we
now calculate the entropy by letting $r_1=b$ and $r_2=b+
\epsilon$:
\begin{equation}
 S = 4\pi\int _{b}^{b+\epsilon} s r^2\sqrt{e^{\lambda}}dr.
\end{equation}
Here $s$ is the entropy density, which may be written as
\begin{equation}
 \frac{\alpha^2k_B^2T(r)}{4\pi\hbar^2G}
= \alpha\left(\frac{k_B}{\hbar}\right)\sqrt{\frac{p}{2 \pi G}},
\end{equation}
where $\alpha^2$ is a dimensionless constant.

Thus the entropy of the fluid within the shell is
\begin{eqnarray}
S =  4\pi\int _{b}^{b+\epsilon}  r^2
\alpha\left(\frac{k_B}{\hbar}\right) \sqrt{\frac{\frac{1}{2r^2}
-\frac{3\widetilde{\psi}_1}{2 r^3}}{16\pi^2G^2}} \sqrt{\frac
{1}{1/2-\widetilde{\psi}_1/ r}} dr \nonumber \\ = \frac{\alpha
k_B}{\hbar G} \int _{b}^{b+\epsilon} r \sqrt{\frac{r
-3\widetilde{\psi}_1} {r-2\widetilde{\psi}_1}} dr \nonumber  \\ =
\frac{\alpha k_B}{ \hbar G} \left[
\frac{2r+3\widetilde{\psi}_1}{4} x
-\frac{9\widetilde{\psi}_1^2}{8} \ln \left(2x +2r - 5
\widetilde{\psi}_1\right) \right]_{b}^{b+\epsilon}
\end{eqnarray}
where $x=\left(r^2-5r\widetilde{\psi}_1+6\widetilde{\psi}_1^2
\right)^{1/2}$.  If $\widetilde{\psi}_1\rightarrow 0$
and the thickness of the shell is negligibly small compared to
its position from the center of the gravastar (i.e., if
 $\epsilon << b$), then the entropy is given by
 $S\approx \frac{ \alpha k_B}{\hbar G} b \epsilon$.\\

\section{The unknown constants $\widetilde{\psi}_0$
and $\widetilde{\psi}_1$}
\noindent
In this section we determine the approximate values of the
constants $\widetilde{\psi}_0$ and $\widetilde{\psi}_1$.
To this end we recall that the thin shell of a gravastar
consists of a perfect stiff fluid \cite{Mazur2001}.
Earlier we denoted the outer radius by $r=a$, the junction
surface.  So by Eq. (\ref{E:perfect}), $\rho=p=p_t$.
Furthermore, for $a$ sufficiently large, $p_t=-v\approx
\Sigma$ by Eq. (\ref{E:omega}).  So we have
\begin{eqnarray}
- \frac{1}{4\pi a}\left[ \sqrt{ 1-\frac{2M}{a} +
\frac{Q^2}{a^2}} -{\widetilde{\psi}_0 a}\right] \nonumber
\\\approx\frac{1}{4\pi a}\left[\frac{ 1 -M/a}
{2 \sqrt{ 1-2M/a +
Q^2/a^2}} -{\widetilde{\psi}_0 a}\right] \nonumber
\\ =\frac{1}{8\pi}\left(\frac{1}{2a^2}
-\frac{3\widetilde{\psi}_1}{2a^3}\right).~~~~~~~~~~~~~~~~~~~~~~~~~~~~~~
\end{eqnarray}
These two equations then yield the approximate values
of the unknown constants:
\begin{equation}
\widetilde{\psi}_0\approx
\frac{3-5M/a+2Q^2/a^2}{4a\sqrt{1-2M/a+Q^2/a^2}},
\end{equation}
\begin{equation}
\widetilde{\psi}_1\approx \frac{a}{3} \left( 1+
\frac{a-3M+2Q^2/a^2}{\sqrt{1-2M/a+Q^2/a^2}} \right).
\end{equation}

Since $\Sigma >0$, Eq.~(\ref{Eq39}) also implies
that $\widetilde{\psi}_0
>\frac{1}{a} \sqrt{1-2M/a + Q^2/a^2} $.\\

\section{Conclusions}\noindent
This paper discusses a new model of a gravastar admitting
conformal motion, within the framework of the Mazur-Mottola model.
The gravastar is assumed to be internally charged with an
exterior  defined by a Reissner-Nordstr{\"o}m rather than a
Schwarzschild metric.  The solutions obtained cover
(i) the interior region, (ii) the shell, and (iii) the
exterior region of the sphere.  Of these three cases, the first
case is of particular interest because the total gravitational
mass $m(r=a)=\frac{1}{2\sqrt 2}(3 -8\widetilde{\psi}_0^2q^2)q$,
obtained by  matching the interior solution to the exterior
Reissner-Nordstr{\"o}m solution at the boundary,
vanishing for vanishing charge $q$ and  turning the total
gravitational mass $m$ into an EMM under the constraint
$\widetilde{\psi}_0<\sqrt 3/(2\sqrt{2})q$. This, in turn, suggests
that the interior de Sitter void of a charged gravastar, having the
same form as the EMM model, must generate the gravitational mass
that provides the attractive force resulting from the collapse of
the sphere and which counter-balances the repulsive force due to
the charge. Moreover, the equation of state $p=-\rho$, known in
the literature as a false vacuum or a $\rho$-vacuum, suggests  that
the {\it charged gravastar} is connected with the {\it dark
star}.

An analysis of the stress energy tensor of the thin shell has
shown that given the requirement of a positive density and positive
pressure, the equation of state parameter $\omega(a)$ is always
positive. Moreover, as the radius of the thin shell increases,
$w(a) \rightarrow 1$.  For some value of $r=a$, we may have
$p_t =0$, yielding a dust shell.

In calculating the  entropy, it was found that if
$\widetilde{\psi}_1\rightarrow 0$ and if, in addition,  the
thickness of the shell is small compared to the radius, then
the entropy is given by
$S\approx \frac{ \alpha k_B}{\hbar G} b \epsilon$, where $b$
is the inner radius of the gravastar shell.  The final
calculations determined the approximate values of the constants
$\widetilde{\psi}_0$ and $\widetilde{\psi}_1$ with
$\widetilde{\psi}_0 >\frac{1}{a} \sqrt{1-2M/a + Q^2/a^2} $.

\section*{Acknowledgments}
\noindent AAU, FR, SR, and KKN wish to thank the authorities of
the Inter-University Centre for Astronomy and Astrophysics, Pune,
India, for providing the Visiting Associateship under which a
part of this work was carried out. FR and ZH are also grateful to
PURSE, Govt. of India and D.S. Kothari fellowship,
 for financial
support. We are also grateful to Prof. Andrew DeBenedictis
 for several
enlightening discussions.

\end{document}